# Automotion of Domain Walls for Spintronic Interconnects

Dmitri E. Nikonov, Sasikanth Manipatruni, Ian A. Young

Exploratory Integrated Circuits, Components Research, Intel Corp., Hillsboro, Oregon 97124, USA

## Abstract

We simulate automotion, the spontaneous transport of a magnetic domain wall under the influence of demagnetization and magnetic anisotropy, in nanoscale spintronic interconnects. In contrast to spin transfer driven magnetic domain wall motion, the proposed interconnects operate with only a transient current pulse and provide favorable scaling down to the 20nm scale. Cases of both in-plane and perpendicular magnetization are considered. Analytical dependence of the velocity of domain walls on the angle of magnetization are compared with full micromagnetic simulations. Deceleration, disappearance, and reflection of domain walls are demonstrated. Dependences of the magnetization angle on the current pulse parameters are studied. The energy and delay analysis suggests that automotion is an attractive option for spintronic logic interconnects.

The development and Moore's law [1] scaling of complementary metal-oxide-semiconductor (CMOS) field effect transistor (FET) electronics over the last four decades was tremendously successful. In recent years, active research has been underway to find viable devices which will supplement CMOS. Currently, many beyond CMOS options are considered [2,3] and many of them are spintronic devices [4], i.e., ones based on spin or magnetic moment as a computational variable. Among them, some are based on motion of domain walls (DW) [5,6,7,8] in ferromagnetic (FM) wires. In general a logic technology comprises of switches and memory elements, interconnected by a complex network of interconnects. Interconnects play a major role in the total power and performance of a computing device. Hence, it is of great interest to identify interconnect technologies for spin based computing. Domain wall spin interconnects avoid spin to electrical conversion and are one of the natural choices for spin based computer. Though DW can be moved over significant distances by a magnetic field [9], driving them with current proved more efficient and convenient [10]. It was treated theoretically [11,12] and observed [13]. The current in plane (CIP) flows, along the FM wire, and the spin transfer torque caused by electron spins' flipping moves the DW. An alternative way of motion with the current perpendicular to plane (CPP) of the FM wire was proposed [14]. It has the obvious disadvantage – the current needs to be applied over the whole length of the FM wire. Most recently, DW motion by the torque of the spin Hall effect was proposed [15]. In that case, the CIP flows in parallel to the FM wire, though the spin torque is applied perpendicularly to it. Traditionally, walls separating domains with magnetization in plane of the chip are considered. Later perpendicular magnetization materials became available, and it was found that DW in them can be moved by a smaller current [16]. However, the existing domain wall based logic and interconnects assume a constant driving current for motion of the domain wall. Combined with the limited efficiency of spin torque and the resistivity of magnetic interconnects, existing DW devices suffer from large energy-delay product [3].

In all options above, a driving force – a magnetic field or a current – was needed for the motion of DW. Only a few works deal with automotion of DW [17], i.e., motion of the wall due to its shape, under the influence of the demagnetization and anisotropy of the FM. Of course, a driving force is needed to create the required initial magnetization distribution in a DW. However, the DW travels a significant distance even after the force is turned off. In [17], the vortex-type DW are formed by CIP. In spite of the attractive energy arguments, automotion gets relatively little attention, even though this regime can be derived from analytical expressions [11] for current-driven motion. The exact dynamics of spontaneous motion of DW also remains to be established. The exact nature of spontaneous motion of DW is described contradictorily. Some works state that "In the absence of the external field, the DW moves back to its original location after the current is turned off" [18]. Conversely other works do predict transient displacement of DW after finite pulses of current [19].

In this letter, we propose the use of automotion of domain walls for local spintronic interconnects. We provide analytical treatment and numerical simulations of automotion of DW to prove their suitability for that purpose. We consider the process of their formation by the spin torque from a CPP pulse and determine the DW's angle and velocity. We show how DW decelerate due to damping and how they can disappear off the edge of a FM wire reflect from it. Finally we estimate the delay and the required energy of an interconnect.

We start by deriving the dynamics of spontaneous motion of DW after a transient spin torque pulse creates it. The dynamics of magnetization in the DW motion is described using the Landau-Lifshitz-Gilbert (LLG) equation [20,21]

$$\frac{dm}{dt} = \alpha \left[ m \times \frac{dm}{dt} \right] - \gamma \left[ m \times B_{eff} \right] + \Gamma_{STT}, \qquad (1)$$

In which the CPP spin torque term is

$$\Gamma_{STT} = -\gamma b_J \left[ m \times p \right] - \gamma a_J \left[ m \times \left[ m \times p \right] \right]. \qquad (2)$$

The unit vector of magnetization is $m$ and its magnitude is $M_s$, the unit vector of injected spin polarization is $p$, the Gilbert damping coefficient is $\alpha$, the Lande g-factor is $g$, the gyromagnetic coefficient is

$$\gamma = \frac{|g|\mu_B}{\hbar}. \qquad (3)$$

The Slonczewski torque and filed-like torque terms are $a_J$ and $b_J \sim 0.1 a_J$,

$$a_J = \frac{\hbar P J}{2 d e M_s}, \qquad (4)$$

the layer thickness is $d$. The effective magnetic field is proportional to the gradient of the total energy of the magnet relative to magnetization

$$B_{eff} = -\frac{1}{M_s} \frac{\delta U}{\delta m}. \qquad (5)$$

The energy per unit volume has the terms for the Zeeman energy in the external field $H$, the exchange energy with a constant $A$, and the combined demagnetization and material anisotropy terms:

$$U = -\mu_0 M_s m \cdot H + A(\nabla m)^2 + K_x m_x^2 + K_y m_y^2 + K_z m_z^2 \tag{6}$$

Where the energy associated with each axis consists of its demagnetization (aka shape anisotropy) with the constants $(N_x, N_y, N_z)$ and material anisotropy with constants $(K_{m,x}, K_{m,y}, K_{m,z})$. The sum of these two parts of energy is overall anisotropy.

We do a numerical solution of the above model using the NIST's OOMMF simulator [22]. We consider a FM wire of length $l = 600nm$ along x-axis, width $w = 20nm$ along y-axis, and thickness $d = 2nm$ along z-axis. Typical local interconnect length needs to be [3] $L_{int} = 20w = 400nm$.

We start with DW in a FM wire with perpendicular magnetization. In an analytical solution for a DW, we neglect the magnetization variations across the wire, and only consider variations in time and along the wire, x-axis, $m(x,t)$. For convenience we represent magnetization by its spherical angles: the polar angle $\theta$ relative to z-axis and the azimuthal angle $\varphi$ in the xy-plane. We arrive at simplified equations similar to those in [18,14] but no driving force, i.e., external field or spin torque

$$\frac{\partial \theta}{\partial t} + \alpha \sin\theta \frac{\partial \varphi}{\partial t} = \frac{\gamma}{M_s}(K_x - K_y)\sin 2\varphi \sin\theta, \tag{7}$$

$$\sin\theta \frac{\partial \varphi}{\partial t} - \alpha \frac{\partial \theta}{\partial t} = \frac{\gamma}{M_s}\left(K_x \cos^2 \varphi + K_y \sin^2 \varphi - K_z\right)\sin 2\theta - \frac{2\gamma A}{M_s}\left(\frac{\partial^2 \theta}{\partial x^2}\right), \tag{8}$$

Then we substitute a Walker trial function [9] describing a decelerating Neel DW

$$\theta(x,t) = 2\arctan\left(\exp\left(\frac{x - u(t)t}{\Delta(t)}\right)\right), \tag{9}$$

which cancels the right hand side of (8) provided that $\Delta = \sqrt{A/K_{eff}}$. DW with an opposite direction of magnetizations correspond to an opposite sign of the inner bracket in (9). For perpendicular magnetization, the effective anisotropy is

$$K_{eff,perp} = K_x \cos^2 \varphi + K_y \sin^2 \varphi - K_z. \tag{10}$$

We assume that the DW angle $\varphi(t)$, width $\Delta(t)$, and velocity $u(t)$ can be functions of time but not of coordinates in the frame tied to the DW. The set of equations turns to

$$-\frac{\Delta}{\alpha}\frac{\partial \varphi}{\partial t} = u_{max}\sin 2\varphi = u + t\frac{du}{dt}, \tag{11}$$

where the maximal velocity of DW is

$$u_{max,perp} = \frac{\gamma \Delta}{M_s(1+\alpha^2)}\left(K_y - K_x\right). \tag{12}$$

Initial velocity of DW is determined by its angle as per (11). The case of in-plane magnetization is treated similarly with the polar angle measured from the x-axis and with the modified expression for the effective anisotropy:

$$K_{eff,inp} = K_y \cos^2 \varphi + K_z \sin^2 \varphi - K_x. \tag{13}$$

The maximum velocity of in-plane DW is

$$u_{max,inp} = \frac{\gamma \Delta}{M_s(1+\alpha^2)} (K_z - K_y). \tag{14}$$

The qualitative behavior described by equations (11) is as follows. The initial azimuthal angle of the DW determines its velocity (both the magnitude and the direction). As noted above, the velocity has the opposite sign for the opposite magnetizations in a DW, as is the case in simulated examples below. Then the azimuthal angle approaches zero (or $\pi$) and the velocity decreases accordingly. Finally a DW stops. This evolution is similar to the Walker breakdown [9]. We show that nanoscale DW automotion interconnects are amenable for a multi-scale interconnect topology (i.e., range of the signal in the interconnect increase with increasing width) required for a micro-chip. We estimate the decay length and decay time of the domain wall automation. An estimate of the time of a DW slowdown based on approximating terms in Eq. (11) is

$$t_{dec} \sim \frac{\Delta}{2\alpha u_{max}} \tag{15}$$

and the distance traveled

$$x_{dec} \sim \frac{\Delta}{2\alpha}. \tag{16}$$

From this, in order to increase the distance of automotion, one needs to increase the DW width (along with the obvious way of decreasing damping). This behavior is illustrated by simulations in Figs. 1 and 2. In all simulations in this letter, a CPP torque is applied to an area of 20x20nm on the left if the FM wire for a certain duration and then switched off. In the patterns of magnetization, the projections of magnetization in-plane are shown by arrows, and the out-of-plane projection corresponds to color: red = positive, and blue = negative. For these simulations we choose the exchange constant $A = 2 \cdot 10^{-11} J/m$, polarization $P = 0.9$, field like torque $b_J = 0.3 a_J$, and Gilbert damping $\alpha = 0.01$, unless stated otherwise. For in-plane magnetization (Fig. 1) we take typical material parameters: $M_s = 1 MA/m$, $K_{m,z} = 0$. This results in DW parameters: $\Delta_{inp} = 8nm$, $u_{max,inp} = 671 m/s$, $x_{dec} = 400nm$, $t_{dec} = 0.6ns$. For the perpendicular magnetization (Fig. 2) we take $M_s = 0.4 MA/m$, $K_{m,z} = 1.2 \cdot 10^5 J/m^3$. This results in DW parameters: $\Delta_{perp} = 23nm$, $u_{max,perp} = 118 m/s$, $x_{dec} = 1144nm$, $t_{dec} = 9.7ns$. The domain walls with in-plane magnetization have a higher velocity mainly due to a larger difference of energies between axes in (14) and also due to a larger magnetization value we used. We see that as the current is applied, a DW is formed and its angle changes. Then a DW angle decreases and it stops. The simulations agree well with the above estimates.

The cases when a DW makes it to the other end of a FM wire are shown in Figs. 3 and 4. The in-plane DW disappears off the end (Fig. 3, between snapshots #6 and #7). After that oscillating and decaying spin waves are radiated along the FM wire. This is seen from a sequence of magnetization arrows deflected in opposite y-axis directions and red and blue regions corresponding to opposite z-axis projections. The perpendicular DW is reflected off the edge and continues propagating back along the FM wire. This difference of behavior is explained by a different character of the demagnetization field generated by magnetic poles at the end of the wire. In the in-plane case, the poles are determined by the magnetization on one side of the DW and exist from the beginning. Their demagnetization field rotates magnetization in the DW so as to promote its continued motion off the edge. In the perpendicular case, the poles are formed by the x-projection of magnetization in the approaching DW. Their demagnetization field rotates magnetization in the DW so as to oppose its motion. Therefore the DW angle changes to opposite and a DW continues motion in the opposite direction. For a spintronic interconnect, reflection of DW is an undesirable feature, since one requires switched magnetization at the end.

Next we study the dependence of DW motion on the spin torque parameters. For that we record the average magnetization in the FM wire and calculate the DW position and angle via the ratio of magnetization projections. For in-plane magnetization (Fig. 5) we see a gradual change of the angle as a function of current, except for a few points when the angle jumps by $\pi$. That happens at a boundary of the range where a DW with a negative velocity is formed and immediately disappears of the left edge (designated as zero velocity and zero angle here). One can see several periods of the angle change over the simulated range of current in Fig. 5. The relation (11) between the angle and the velocity is confirmed by simulation with an accuracy of ~10%. The reasons for discrepancy are the approximations of the analytical solution, the discretization errors of the numerical solution, and slight oscillations of the DW angle found in the simulation. The dependence of the DW velocity and angle on the pulse duration (Figs. 6 and 8) is much more flat. At a short pulse duration, spin torque is not sufficient to flip magnetization and thus to create a DW. At long pulse duration, the DW is already travelling away from the area of spin torque, and the spin torque does not affect its parameters.

The dependence of the motion of DW with perpendicular magnetization, shown in Fig. 7, is more oscillatory. The DW velocity turns to zero at separate points at which $\sin 2\varphi$ turns to zero. At ranges of current between such points, a DW may be formed with a negative velocity. In this case a DW is immediately reflected off the left edge and starts propagating right with a positive velocity. Therefore perpendicular domain walls have more values of current with higher velocity, comparable to $u_{max}$, and the interconnect operation is less sensitive to current variations.

Finally we estimate the switching energy necessary to create a domain wall and the delay of propagation in the interconnect of 400nm length. From the above simulation the characteristic values for in-plane DW are $t_{ic} = 0.6ns$, $E_{ic} = 10 fJ$, and for perpendicular DW are $t_{ic} = 3.5ns$, $E_{ic} = 7.8 fJ$. The DW interconnects benefit from a low voltage $V_{dw} = 0.1V$ at which spin torque switching can occur. Therefore in-plane the DW interconnect is much faster at a price of a modestly higher switching energy. These values of energy are better than those with DW persistently driven by spin torque and are competitive with benchmarks of beyond CMOS circuits [3]. On a different metric, energy per bit per unit length, DW interconnects score ~20fJ/bit/µm which is competitive even with CMOS. For the latter we can estimate the switching energy as $E = V_{dd}^2 t_{sw} w^2 / \rho L_{int}$, with voltage $V_{dd} = 1V$, switching time $t_{sw} = 3ps$, resistivity $\rho = 8 \cdot 10^{-8} m \cdot \Omega$ [23,24] projected for w=20nm wide metallic wires. This will result in 94fJ/bit/µm for CMOS interconnects. On the downside, the delay of the DW interconnects is much longer than that of electronic interconnects, as seen above.

We considered the automotion of the domain walls with both in-plane and perpendicular magnetizations walls which were created by an initial pulse of a current. These domain walls decelerate and tend stop, but before that they traverse significant distances. In-plane domain walls disappear off the ends of wires, while perpendicular domain walls reflect from them. The initial velocity of domain walls depends on the angle of magnetization in the domain wall. This angle has a strongly oscillating dependence on the magnitude of the current causing the rotation by the spin torque. This angle is not very sensitive to the duration of the pulse. In summary, such domain walls are suitable for interconnects between spin logic gates. Domain walls with in-plane magnetization are preferable since they move with a higher velocity than domain walls with perpendicular magnetization and require comparable energy to create them.

[1] G. E. Moore, Electronics 19, 114 (1965).

[2] D. Nikonov and I. Young, Proceedings of the International Electron Devices Meeting (IEDM), 25.4.1-4 (2012).

[3] D. E. Nikonov and I. A. Young, Proceedings of IEEE, early access (2013).

[4] D. E. Nikonov and G. I. Bourianoff, J. Supercond. and Novel Magnetism 21, 479-493 (2008).

[5] D. A. Allwood, G. Xiong, C. C. Faulkner, D. Atkinson, D. Petit, and R. P. Cowburn, Science, 309, 1688 (2005).

[6] J. A. Currivan, Y. Jang, M. D. Mascaro, M. A. Baldo, and C. A. Ross, IEEE Magn. Lett. 3, 3000104 (2012).

[7] D. E. Nikonov, G. I. Bourianoff, and P. A. Gargini, J. Nanoelectron. Optoelectron., 3, 3 (2008).

[8] D. M. Bromberg, D. H. Morris, L. Pileggi, and J.-G. Zhu, IEEE Trans. Magnet. 48, 3215 (2012).

[9] N. L. Schryer and L. R. Walker, J. Appl. Phys. 45, 5406 (1974).

[10] L. Berger, J. Appl. Phys. 55, 1954 (1984).

[11] Z. Li and S. Zhang, Phys. Rev. Lett., 92, 207203 (2004).

[12] S. Zhang and Z. Li, Phys. Rev. Lett. 93, 127204 (2004).

[13] A. Yamaguchi, T. Ono, S. Nasu, K. Miyake, K. Mibu, and T. Shinjo, Phys. Rev. Lett. 92, 077205(2004).

[14] A.V. Khvalkovskiy, K. A. Zvezdin, Ya. V. Gorbunov, V. Cros, J. Grollier, A. Fert, and A. K. Zvezdin, Phys. Rev. Lett. 102, 067206 (2009).

[15] J. Ryu, K.-J. Lee, and H.-W. Lee, Appl. Phys. Lett. 102, 172404 (2013).

[16] S. Fukami, T. Suzuki, N. Ohshima, K. Nagahara, and N. Ishiwata, Appl. Phys. Lett. 103, 07E718 (2008).

[17] J.-Y. Chauleau, R. Weil, A. Thiaville, and J. Miltat, Phys. Rev. B 82, 214414 (2010).

[18] Z. Li and S. Zhang, Phys. Rev. B 70, 024417 (2004).

[19] A. Thiaville, Y. Nakatani, F. Pi´echon, J. Miltat, and T. Ono, Eur. Phys. J. B 60, 15–27 (2007).

[20] D.C. Ralph, M.D. Stiles, J. Magn. Magn. Mater. 320, 1190 (2008).

[21] D. V. Berkov and J. Miltat, J. Magn. Magn. Mater. 320, 1238 (2008).

[22] M. J. Donahue and D. G. Porter, "OOMMF User's Guide, Version 1.0," National Institute of Standards and Technology Report No. NISTIR 6376, September 1999.

[23] J. S. Chawla, R. Chebiam, R. Akolkar, G. Allen, C. T. Carver, J. S. Clarke, F. Gstrein, M. Harmes, T. Indukuri, C. Jezewski, B. Krist, H. Lang, A. Myers, R. Schenker, K. J. Singh, R. Turkot, and H. J. Yoo, IEEE Interconnect Technology Conference (IITC),  p. 1-3, (2013).

[24] R. L. Graham, G. B. Alers, T. Mountsier, N. Shamma, S. Dhuey, S. Cabrini, R. H. Geiss, D. T. Read, and S. Peddeti, Appl. Phys. Lett. 96, 042116 (2010).

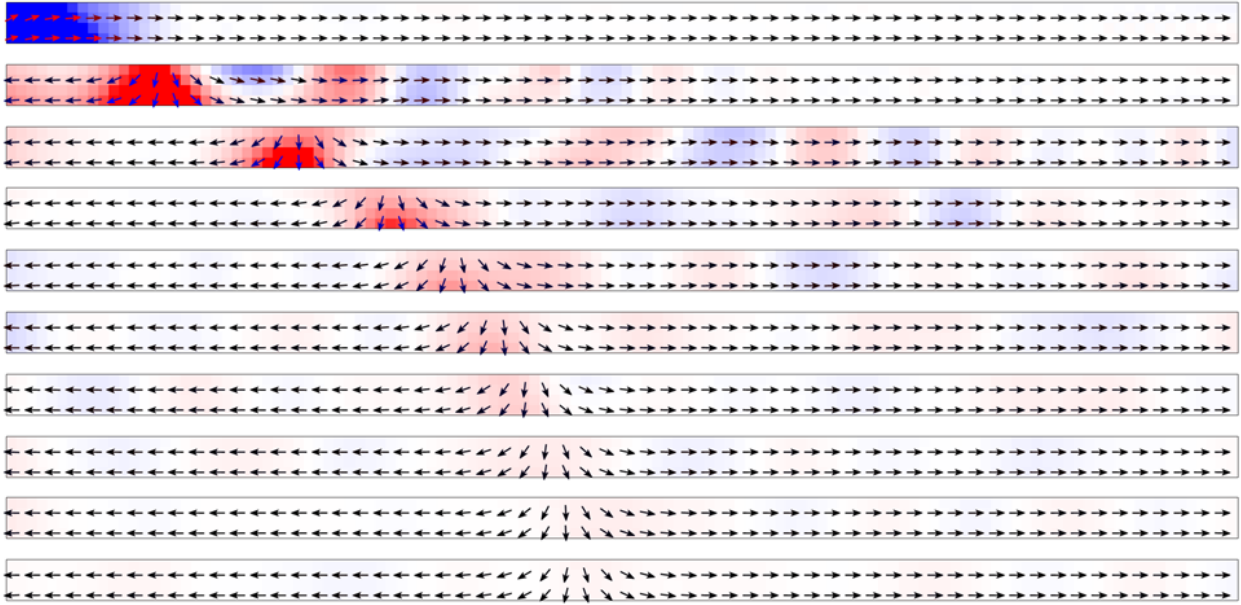

Figure 1. Snapshots of magnetization at intervals of 0.2ns for in-plane DW. Current $I = 300\mu A$, pulse $\tau = 0.5ns$, damping $\alpha = 0.01$.

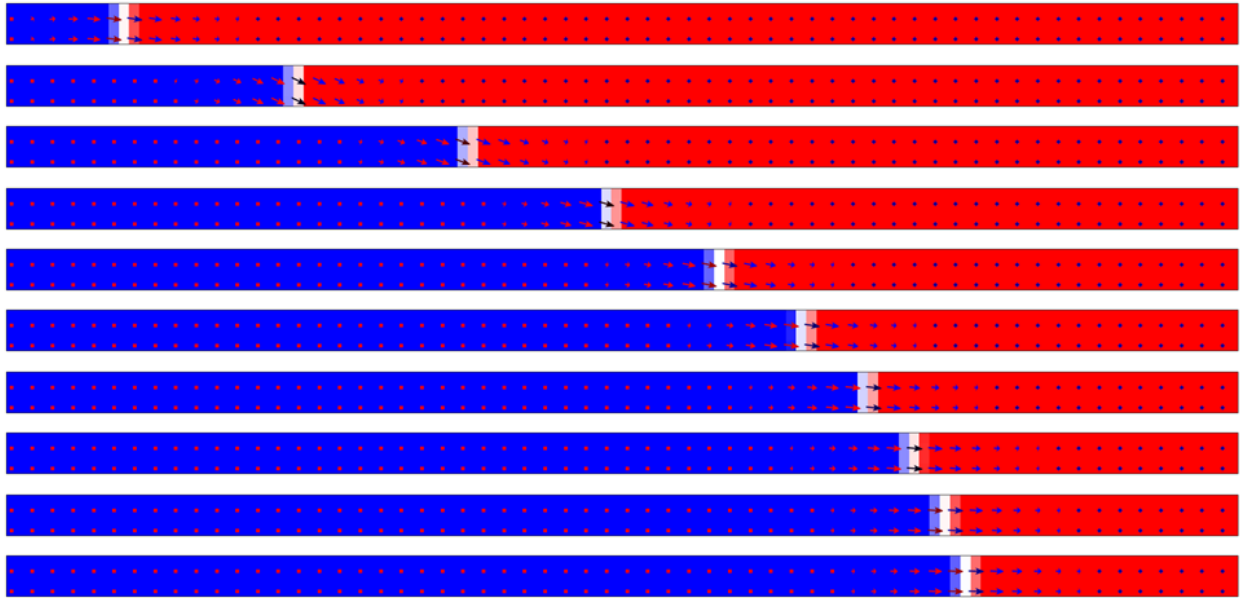

Figure 2. Snapshots of magnetization at intervals of 1ns for perpendicular DW. Current $I = 240\mu A$, pulse $\tau = 2ns$, damping $\alpha = 0.03$.

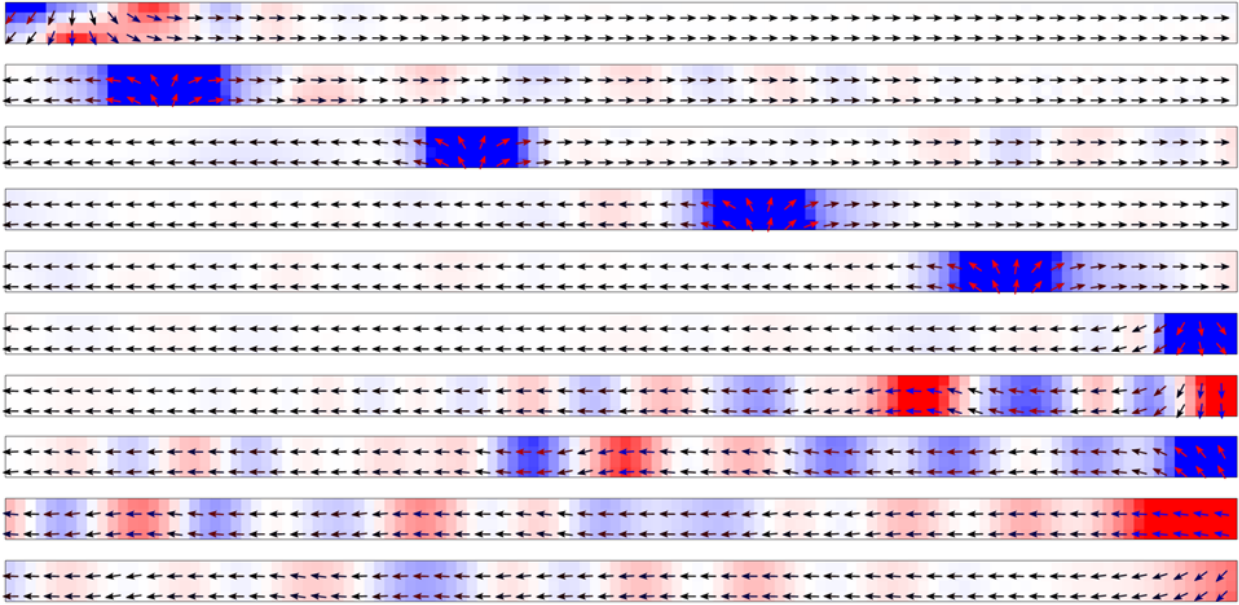

Figure 3. Snapshots of magnetization at intervals of 0.2ns for in-plane DW. Current $I = 400\mu A$, pulse $\tau = 0.5ns$.

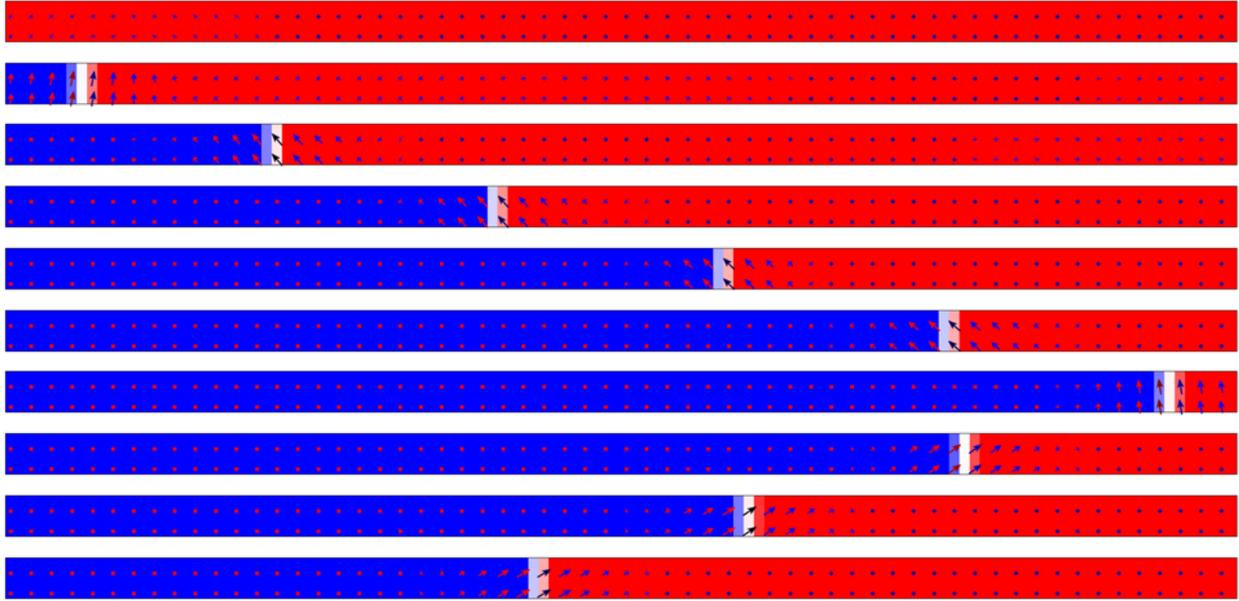

Figure 4. Snapshots of magnetization at intervals of 1ns for perpendicular DW. Current $I = 100\mu A$, pulse $\tau = 2ns$, damping $\alpha = 0.01$.

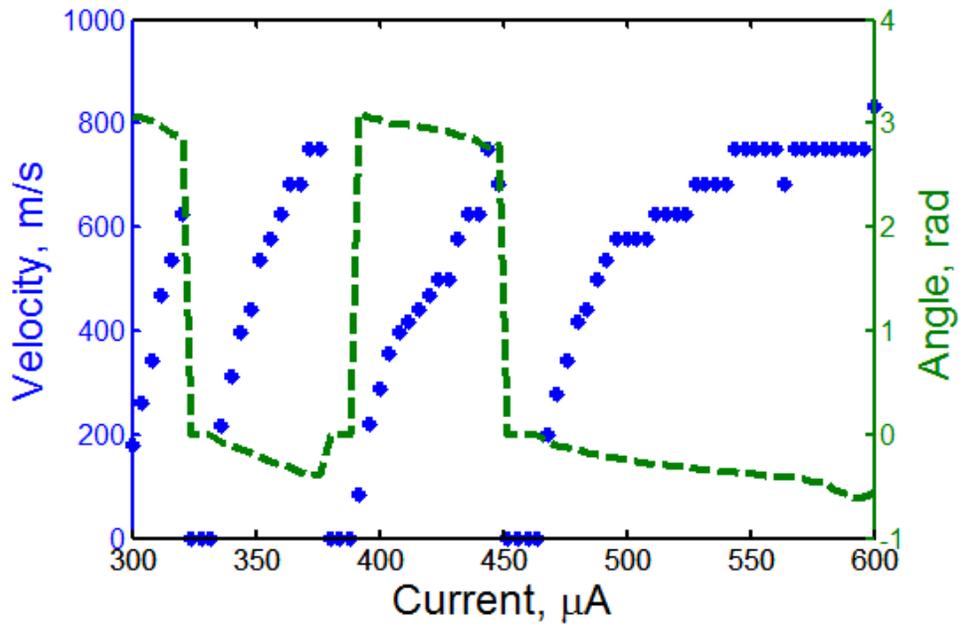

Figure 5. DW velocity and angle vs. current for in-plane magnetization, pulse $\tau = 0.5 ns$. The velocity is designated by unconnected stars, and the angle – by a dashed line.

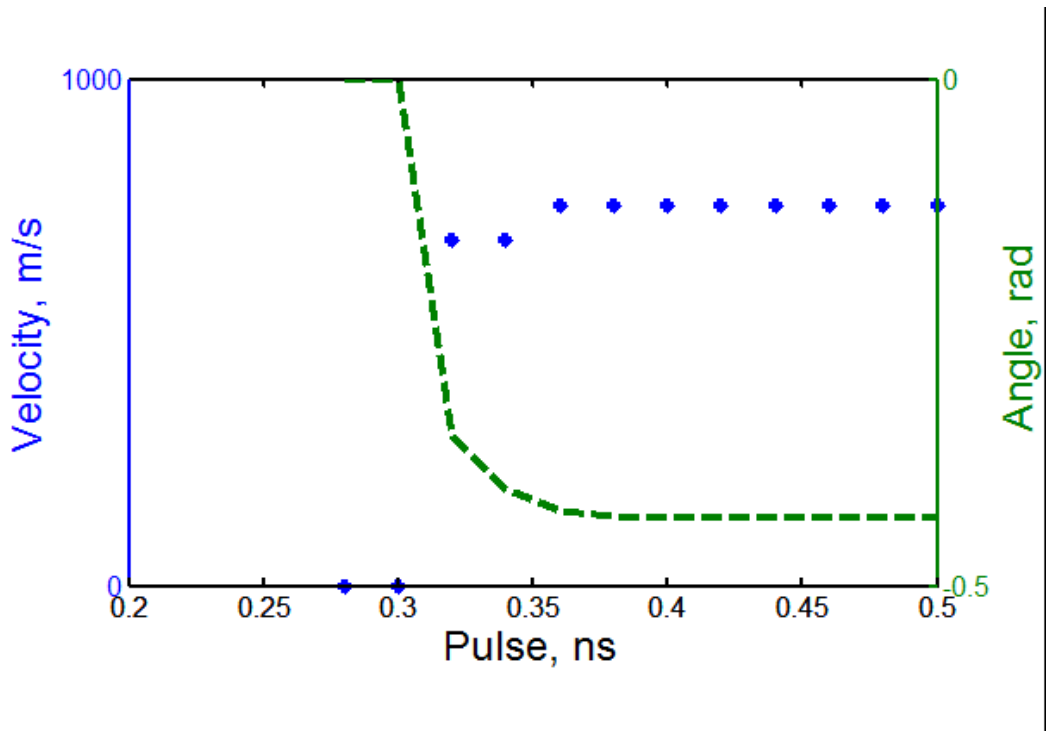

Figure 6. DW velocity and angle vs. pulse duration for in-plane magnetization, current $I = 400 \mu A$.

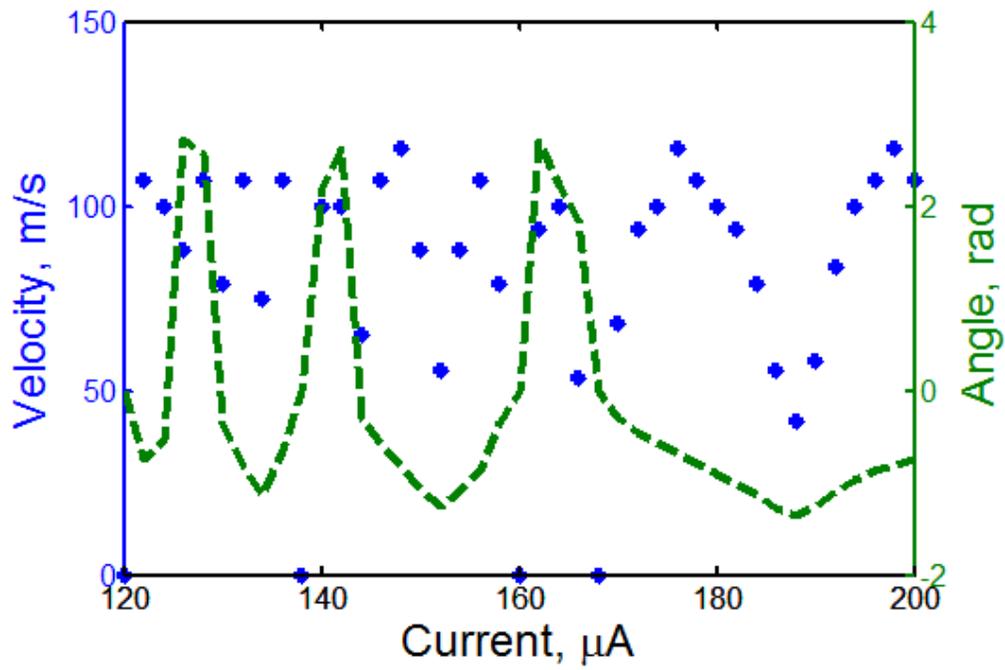

Figure 7. DW velocity and angle vs. current for perpendicular magnetization, pulse $\tau = 1ns$.

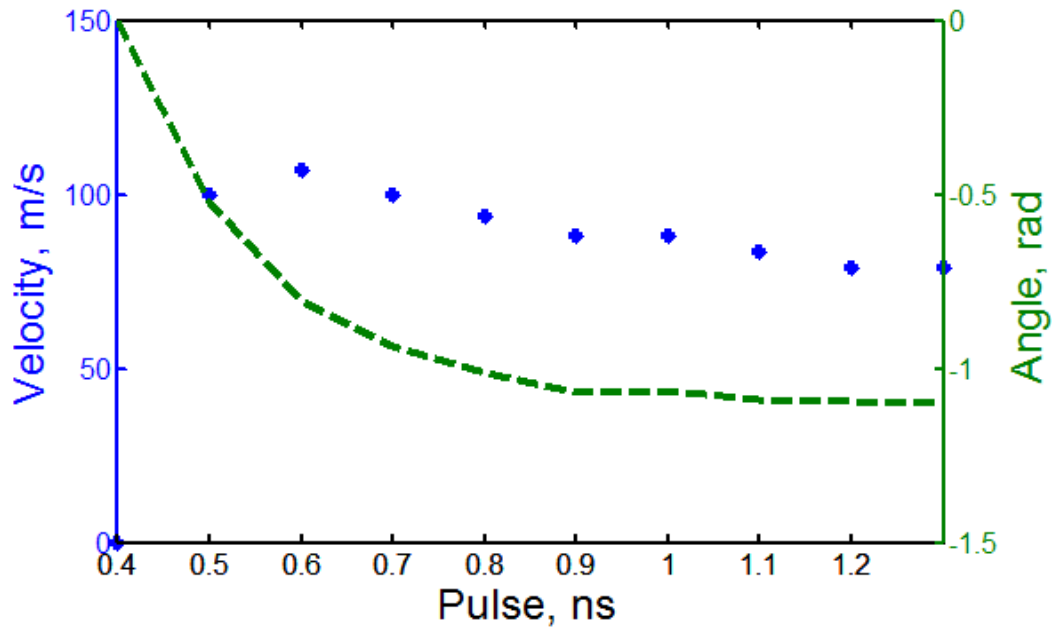

Figure 8. DW velocity and angle vs. pulse duration for perpendicular magnetization, current $I = 150\mu A$.

# Supplementary Material

## Equations for magnetization dynamics

The Landau-Lifshitz-Gilbert (LLG) equations for magnetization are

$$\frac{dm}{dt} = \alpha \left[ m \times \frac{dm}{dt} \right] - \gamma [m \times B_{eff}] + \Gamma_{STT},$$

Where the spin torque term for in-plane current is

$$\Gamma_{STTi} = -(u_J \cdot \nabla)m + \beta m \times [(u_J \cdot \nabla)m],$$

where the velocity corresponding to the in-plane current density $J$ is

$$u_J = \frac{|g|\mu_B PJ}{2eM_s}$$

And for out of plane current the spin torque term is

$$\Gamma_{STTp} = -\gamma b_J [m \times p] - \gamma a_J [m \times [m \times p]].$$

The unit vector of magnetization is $m$, the unit vector of injected spin polarization is $p$, the Gilbert damping coefficient is $\alpha$, the magnetization is $M_s$, the Lande g-factor is $g$, the gyromagnetic coefficient is $\gamma$

$$\gamma = \frac{|g|\mu_B}{\hbar}.$$

The spin torque terms are $b_J \sim 0.1 a_J$, the layer thickness is $d$, and

$$a_J = \frac{\hbar PJ}{2deM_s}$$

The effective magnetic field is proportional to the gradient of the total energy of the magnet relative to magnetization

$$B_{eff} = -\frac{1}{M_s} \frac{\delta U}{\delta m}.$$

The energy per unit volume has the terms for the Zeeman energy in the external field $H$, the exchange energy with a constant $A$, and the combined demagnetization and material anisotropy terms:

$$U = -\mu_0 M_s m \cdot H + A(\nabla m)^2 + K_x m_x^2 + K_y m_y^2 + K_z m_z^2$$

Where the energy associated with each axis consists of its demagnetization (aka shape anisotropy) with the constants $(N_x, N_y, N_z)$ and material anisotropy with constants $(K_{m,x}, K_{m,y}, K_{m,z})$. The sum of these two parts of energy is overall anisotropy

$$K_x = \frac{1}{2}\mu_0 M_s^2 N_x + K_{m,x},$$

and similarly for the other two axes.

Performing the gradient we obtain that the effective field in vector form

$$B_{eff} = \mu_0 H - \frac{2A}{M_s}\nabla^2 m - \frac{2}{M_s}\mathbf{K} \cdot m$$

### Simplification for domain walls in a magnetic wire
In an analytical solution for a domain wall, we neglect the magnetization variations across the wire, and only consider variations in time and along the wire, x-axis, $m(x,t)$.

For convenience we represent magnetization by its spherical angles: $\theta$ relative to z-axis and $\varphi$ in the xy-plane of the chip. It is especially convenient for wires with equilibrium **magnetization out-of-plane** of the chip (in-plane magnetization is treated similarly):

$$m_z = \cos\theta$$
$$m_x = \sin\theta\cos\varphi$$
$$m_y = \sin\theta\sin\varphi$$

Their projections on spherical coordinates

$$m_z = [-\sin\theta\cos\theta, 0]$$
$$m_x = \sin\theta\cos\varphi[\cos\theta\cos\varphi, -\sin\varphi]$$
$$m_y = \sin\theta\sin\varphi[\cos\theta\sin\varphi, -\sin\varphi]$$

In the spherical coordinates, a differential is

$$dm = [d\theta, \sin\theta d\varphi]$$

The Laplacian is

$$\frac{\partial^2 m}{\partial x^2} = \left[\frac{\partial^2 \theta}{\partial x^2} - \sin\theta\cos\theta\left(\frac{\partial \varphi}{\partial x}\right)^2, 2\cos\theta\frac{\partial \theta}{\partial x}\frac{\partial \varphi}{\partial x} + \sin\theta\frac{\partial^2 \varphi}{\partial x^2}\right]$$

For any vector in $v = [v_\theta, v_\varphi]$ spherical coordinates, the cross product is

$$[m \times v] = [-v_\varphi, v_\theta]$$

The cross-product of unit vectors along axes with the magnetization vector is

$$[m \times \hat{z}] = [0, -\sin\theta]$$
$$[m \times \hat{x}] = [\sin\varphi, \cos\theta\cos\varphi]$$
$$[m \times \hat{y}] = [\sin\varphi, \cos\theta\sin\varphi]$$

Using these expressions we can re-cast the terms of the LLG equation from Cartesian to spherical coordinates.

Then projecting the LLG equations on the local spherical coordinates

$$\frac{\partial \theta}{\partial t} + \alpha\sin\theta\frac{\partial \varphi}{\partial t} = -\gamma a_J \sin\theta + \mu_0\gamma\left(-H_x\sin\varphi + H_y\cos\varphi\right)$$

$$+ \frac{\gamma}{M_s}(K_x - K_y)\sin 2\varphi \sin\theta + \frac{2\gamma A}{M_s}\left(2\cos\theta\frac{\partial \theta}{\partial x}\frac{\partial \varphi}{\partial x} + \sin\theta\frac{\partial^2 \varphi}{\partial x^2}\right)$$

$$\sin\theta\frac{\partial \varphi}{\partial t} - \alpha\frac{\partial \theta}{\partial t} = \gamma b_J \sin\theta + \mu_0\gamma\left(H_z\sin\theta - H_x\cos\theta\cos\varphi - H_y\cos\theta\sin\varphi\right)$$

$$+ \frac{\gamma}{M_s}\left(K_x\cos^2\varphi + K_y\sin^2\varphi - K_z\right)\sin 2\theta + \frac{2\gamma A}{M_s}\left(-\frac{\partial^2 \theta}{\partial x^2} + \sin\theta\cos\theta\left(\frac{\partial \varphi}{\partial x}\right)^2\right)$$

### Domain wall solutions
If we are aiming to find solutions of the domain walls with constant $\varphi$ and constant velocity of propagation, we can substitute the functional shape for static domain walls

$$\theta(x,t) = 2\arctan\left(\exp\left(\frac{x - ut}{\Delta}\right)\right)$$

It is remarkable for the fact that

$$\frac{d\theta}{\sin\theta} = d\left(\frac{x-ut}{\Delta}\right)$$

Or

$$\frac{1}{\sin\theta}\frac{d\theta}{dt} = \frac{-u}{\Delta}$$

Here is the width of a domain wall to be determined from the equations.

Then the equations turn to

$$-\frac{u}{\Delta}\sin\theta = -\gamma a_J \sin\theta + \mu_0\gamma\left(-H_x \sin\varphi + H_y \cos\varphi\right)$$
$$+\frac{\gamma}{M_s}\left(K_x - K_y\right)\sin 2\varphi \sin\theta$$

$$\frac{\alpha u}{\Delta}\sin\theta = \gamma b_J \sin\theta + \mu_0\gamma\left(H_z \sin\theta - H_x \cos\theta \cos\varphi - H_y \cos\theta \sin\varphi\right)$$
$$+\frac{\gamma}{M_s}\left(K_x \cos^2\varphi + K_y \sin^2\varphi - K_z\right)\sin 2\theta - \frac{2\gamma A}{M_s}\frac{\partial^2\theta}{\partial x^2}$$

In the absence of the spin torque or external magnetic fields the equations further simplify to

$$\frac{u}{\Delta} = \frac{\gamma}{M_s}\left(K_y - K_x\right)\sin 2\varphi$$

$$\frac{\alpha u}{\Delta}\sin\theta = \frac{\gamma}{M_s}\left(K_x \cos^2\varphi + K_y \sin^2\varphi - K_z\right)\sin 2\theta - \frac{2\gamma A}{M_s}\left(\frac{\partial^2\theta}{\partial x^2}\right)$$

From these equations we can approximately find the width and velocity of the domain walls. Substituting the velocity one obtains

$$-\alpha\left(K_y - K_x\right)\sin 2\varphi \sin\theta + \left(K_x \cos^2\varphi + K_y \sin^2\varphi - K_z\right)2\sin\theta\cos\theta = 2A\left(\frac{\partial^2\theta}{\partial x^2}\right)$$

Let us introduce the effective anisotropy energy which is pertinent to the domain wall width

$$K_{eff,perp} = K_x \cos^2\varphi + K_y \sin^2\varphi - K_z,$$

which is positive in the case of $K_y > K_x > K_z$ typical for perpendicular magnetization.

**Neglecting the term with the factor of Gilbert damping $\alpha$ in front of it**, we arrive at the equation

$$K_{eff} 2\sin\theta\cos\theta = 2A\left(\frac{\partial^2\theta}{\partial x^2}\right)$$

Which has the solution

$$\theta(x,t) = 2\arctan\left(\exp\left(\frac{x-ut}{\Delta}\right)\right)$$

with $\Delta = \sqrt{A/K_{eff}}$. This verifies the assumption of the functional shape made above.

Finally the velocity of domain walls is principally determined by their angle $\varphi$

$$u_{perp} = \frac{\gamma\Delta}{M_s}(K_y - K_x)\sin 2\varphi.$$

It can be positive or negative depending on the angle. Its magnitude is maximal for the angle of 45, -45, 135 and -135degrees.

$$u_{max} = \frac{\gamma\Delta}{M_s}(K_y - K_x)$$

The case of in-plane magnetization is treated similarly. The only change is that the spherical angles are now measured from the x-axis rather than the z-axis. Then we can re-use the above results with a cyclic permutation of the x,y,z-indices.

$$K_{eff,inp} = K_y \cos^2\varphi + K_z \sin^2\varphi - K_x,$$

which is positive in the case of $K_z > K_y > K_x$ typical for in-plane magnetization. The velocity of domain walls is

$$u_{inp} = \frac{\gamma\Delta}{M_s}(K_z - K_y)\sin 2\varphi$$

Which may be faster than that for perpendicular polarization.

## Comparison with published results

Let us compare this with analytical equations from A.V. Khvalkovskiy et al., PRL 102, 067206 (2009) for zero external magnetic field and neglecting exchange in the first equation:

$$-\frac{u}{\Delta} + \alpha \frac{d\varphi}{dt} = -\gamma a_J + \frac{\gamma}{M_s}(K_y - K_z)\sin 2\varphi$$

$$\frac{d\varphi}{dt} + \alpha \frac{u}{\Delta} = \gamma b_J$$

Separating the equality for the domain wall width from the second equation

$$(K_y \cos^2 \varphi + K_z \sin^2 \varphi - K_x)\sin 2\theta = 2A\left(\frac{\partial^2 \theta}{\partial x^2}\right),$$

which gives the same expression for the domain wall width as above.

With this treatment one can easily come to a conclusion that steady motion of domain walls with constant angle $\varphi$ is impossible without persistent current to compensate for damping. The corresponding velocity is

$$u = \frac{\Delta \gamma b_J}{\alpha} = 0.1 \frac{\Delta}{d} \frac{\gamma \hbar PJ}{2\alpha e M_s} = 0.1 \frac{\Delta}{d} \frac{g\mu_B PJ}{2\alpha e M_s}$$

The expression is similar (apart from a geometry factor) to the velocity of domain walls driven by in-plane current.

In contrast we are focusing on sufficiently long movement of domain walls without a persistent current.

## Energy dissipation

In fact, damping does sap energy from domain wall motion.

The change of energy with time is

$$\frac{dU}{dt} = \frac{\delta U}{\delta m} \cdot \frac{dm}{dt} = -M_s B_{eff} \cdot \frac{dm}{dt}$$

The terms in LLG corresponding to the effective field are orthogonal to the effective field and thus do not change the energy. This is understandable since these are terms conserving energy. In the absence of spin torque, the only term which dissipates energy is Gilbert damping

$$\frac{dU}{dt} = -\alpha M_s B_{eff} \cdot \left[m \times \frac{dm}{dt}\right]$$

In spherical coordinates it reduces to

$$\frac{dU}{dt} = -\alpha M_s \left( B_{eff,\varphi} \frac{dm}{dt}\bigg|_\theta - B_{eff,\theta} \frac{dm}{dt}\bigg|_\varphi \right)$$

In other words

$$\frac{dU}{dt} = -\alpha M_s \left( \begin{pmatrix} -\frac{2A}{M_s}\left(\frac{\partial^2 \theta}{\partial x^2} - \sin\theta\cos\theta\left(\frac{\partial \varphi}{\partial x}\right)^2\right) \\ -\frac{2}{M_s}\sin\theta\cos\theta(K_x\cos^2\varphi + K_y\sin^2\varphi - K_z) \end{pmatrix} \frac{d\theta}{dt} - \begin{pmatrix} -\frac{2A}{M_s}\left(2\cos\theta\frac{\partial\theta}{\partial x}\frac{\partial\varphi}{\partial x} + \sin\theta\frac{\partial^2\varphi}{\partial x^2}\right) \\ \frac{2}{M_s}\sin\theta\sin\varphi(K_x\cos\varphi + K_y\sin\varphi) \end{pmatrix} \sin\theta\frac{d\varphi}{dt} \right)$$

For the case of constant angle $\varphi$, it simplifies to

$$\frac{dU}{dt} = 2\alpha\left(A\frac{\partial^2\theta}{\partial x^2} + \sin\theta\cos\theta K_{eff,perp}\right)\frac{d\theta}{dt}$$

$$\frac{dU}{dt} = -4\alpha\sin^2\theta\cos\theta K_{eff,perp}\frac{u}{\Delta}$$

Which amounts to non-zero energy dissipation proportional to a small Gilbert damping constant. Therefore we hope that the domain wall can travel over a time of hundreds of characteristic precession periods.

### Decelerated domain walls

Let us examine once again the LLG equations with no driving force, i.e. external field or spin torque. **We will no longer neglect damping.** We assume that angle $\varphi(t)$ and velocity $u(t)$ can now be functions of time but not coordinates tied to domain wall

$$\frac{\partial\theta}{\partial t} + \alpha\sin\theta\frac{\partial\varphi}{\partial t} = \frac{\gamma}{M_s}(K_x - K_y)\sin 2\varphi\sin\theta$$

$$\sin\theta\frac{\partial\varphi}{\partial t}-\alpha\frac{\partial\theta}{\partial t}=\frac{\gamma}{M_s}\left(K_x\cos^2\varphi+K_y\sin^2\varphi-K_z\right)\sin 2\theta-\frac{2\gamma A}{M_s}\left(\frac{\partial^2\theta}{\partial x^2}\right)$$

Re-introducing the domain wall width and shape, we arrive at

$$\frac{1}{\sin\theta}\frac{\partial\theta}{\partial t}+\alpha\frac{\partial\varphi}{\partial t}=\frac{\gamma}{M_s}\left(K_x-K_y\right)\sin 2\varphi$$

$$\frac{\partial\varphi}{\partial t}=\alpha\frac{1}{\sin\theta}\frac{\partial\theta}{\partial t}$$

Then substituting the shape function for the decelerating domain wall

$$\theta(x,t)=2\arctan\left(\exp\left(\frac{x-u(t)t}{\Delta}\right)\right)$$

Which obeys the relation

$$\frac{1}{\sin\theta}\frac{\partial\theta}{\partial t}=-\frac{u}{\Delta}-\frac{t}{\Delta}\frac{du}{dt}$$

We arrive at

$$-\frac{\Delta}{\alpha}\frac{\partial\varphi}{\partial t}=u_{max}\sin 2\varphi=u+t\frac{du}{dt}$$

Where the maximal velocity of domain walls is

$$u_{max}=\frac{\gamma\Delta}{M_s(1+\alpha^2)}\left(K_y-K_x\right)$$

These equations allow us to first obtain the change of angle in time, and then determine the domain wall velocity. Their solution is

$$\tan\varphi=\tan\varphi_0\exp\left(-\frac{2\alpha u_{max}t}{\Delta}\right)$$

The equation for velocity can be solved numerically

$$u+t\frac{du}{dt}=u_{max}\sin 2\varphi$$

The qualitative behavior: the angle decreases towards zero, and the velocity decreases. A rough estimate of the time to slow down is

$$t_{dec} \sim \frac{\Delta}{2\alpha u_{max}}$$

And the distance traveled

$$x_{dec} \sim \frac{\Delta}{2\alpha}$$

Which can be around 1000nm for the domain wall width of 20nm.

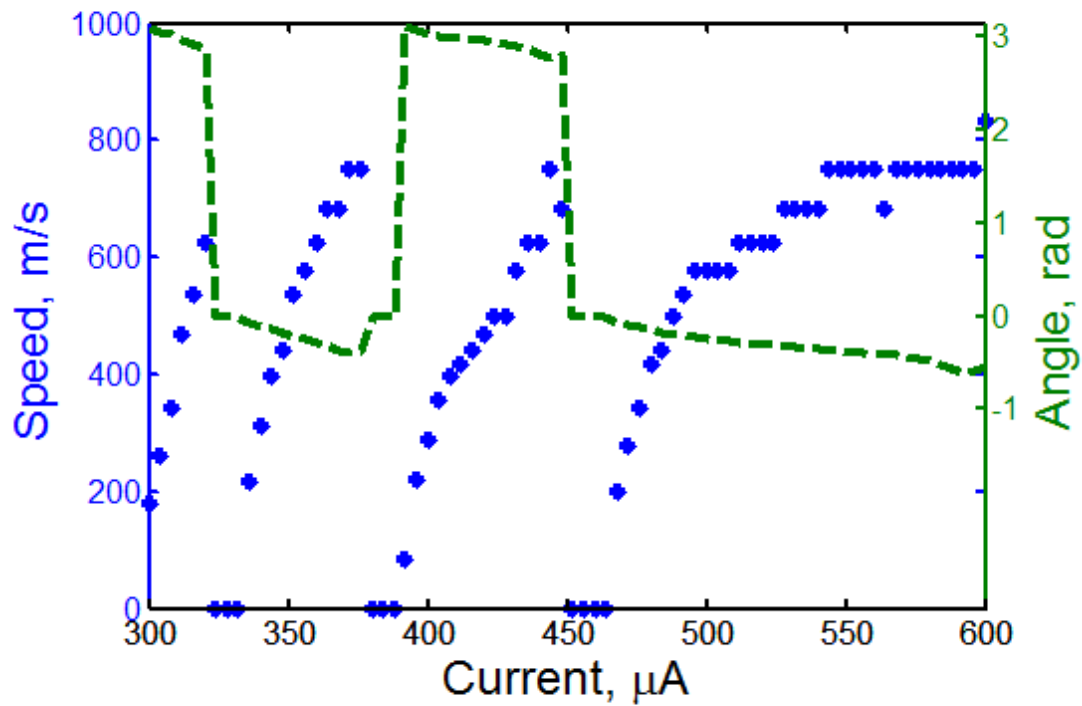

Figure 9. DW velocity and angle vs. current for in-plane magnetization, pulse $\tau = 0.5ns$, with zero field-lie torque.